\documentclass[twocolumn,english,reprint,amsmath,amssymb, onecolumn,nofootinbib,11pt]{revtex4-2}
\usepackage[T1]{fontenc}
\usepackage[latin9]{inputenc}
\makeatletter
\usepackage{soul}
\usepackage{float}
\usepackage{physics}
\usepackage{amsfonts}
\usepackage{latexsym}
\usepackage{epsfig}
\usepackage{graphicx}
\usepackage{tikz}
\definecolor{greatblue}{RGB}{40,120,181}
\definecolor{greatred}{RGB}{200,36,35}
\usepackage[colorlinks,linkcolor=greatblue,anchorcolor=blue,citecolor=greatred]{hyperref}

\usepackage{dcolumn}
\usepackage{bm}

\makeatother

\usepackage[english]{babel}

\usepackage{hyperref}
\usepackage{braket}
\usepackage{booktabs}
\usepackage{array}
\usepackage{paralist} 
\usepackage{verbatim}
\usepackage{fancyhdr}
\begin{document}
\preprint{preprintnumbers}{CTP-SCU/2025002}

\title{Entanglement Entropy of Mixed State in Thermal $\text{CFT}_2$}

\author{Xin Jiang}  
\email{domoki@stu.scu.edu.cn}
\affiliation{College of Physics, Sichuan University, Chengdu, 610065, China}

\author{Haitang Yang}
\email{hyanga@scu.edu.cn}
\affiliation{College of Physics, Sichuan University, Chengdu, 610065, China}

\author{Zilin Zhao}
\email{zhaozilin@stu.scu.edu.cn}   
\affiliation{College of Physics, Sichuan University, Chengdu, 610065, China}

\begin{abstract}
Using the subtraction approach, we give the bipartite mixed state entanglement entropy in thermal $\text{CFT}_2$. With these entanglement entropies, we examine in detail the holographic duals of different entangling configurations unambiguously.
In the thermofield double state, we show a horizon-crossing feature in two-sided entanglement configuration.

\end{abstract}
\maketitle
\newpage 
\tableofcontents
	
\section{Introduction}


In the context of the AdS/CFT correspondence \citep{Maldacena:1997re,Witten:1998qj,Gubser:1998bc}, entanglement entropy serves as a fundamental concept to provide critical insights into the relation between quantum entanglement and the emergence of spacetime \citep{Ryu:2006bv,Ryu:2006ef,Hubeny:2007xt,VanRaamsdonk:2010pw,Jiang:2024xcy,Jiang:2024xqz}. 
Recent progress    reveals how the Einstein equation emerges from the quantum entanglement in CFT$_2$ \cite{Jiang:2024xcy}. $ER=EPR$ has been also realized with some well-designed setup \cite{Jiang:2024xqz}.

For a pure state $\Sigma=A\cup B$, the total density matrix is $\rho=|\Sigma\rangle\langle\Sigma|$. 
The entanglement entropy between the two complementary subsystems $A$ and $B$ is defined by the von Neumann entropy,
\begin{equation}
S_{\mathrm{vN}}\left(\rho_{A}\right)=-\mathrm{Tr}_{A}\rho_{A}\log\rho_{A},
\end{equation}
where the reduced density matrix $\rho_{A}$  is  $\rho_A = \mathrm{Tr}_{B}\rho$. 

However, when dealing with bipartite systems in a mixed state, such as two disjoint intervals $A$ and $B$ in CFT$_2$, a straightforward generalization of the definition above proves to be an inaccurate measure for quantifying quantum entanglement between $A$ and $B$. 
To address this issue, several candidates have been proposed to quantify the quantum entanglement in mixed states, 
such as entanglement negativity \citep{Calabrese:2012nk},  entanglement of purification \cite{Takayanagi:2017knl}, odd entanglement entropy \citep{Tamaoka:2018ned}, reflected entropy \citep{Dutta:2019gen}, balanced partial entanglement \citep{Wen:2021qgx} and the distillable entanglement \citep{Mori:2024gwe}.

The entanglement of purification (EoP) \cite{Takayanagi:2017knl} is a well studied correlation among the proposed quantities. The basic idea of EoP is to purify the mixed state by enlarging the Hilbert space of the mixed state with auxiliary systems. For a mixed state $\rho_{AB}$, one introduces auxiliary states $A^*$ and $B^*$ to form a pure state $\rho_{AA^* BB^*}$ under the condition $\rho_{AB} = \mathrm{Tr}_{A^* B^*} \rho_{AA^* BB^*}$. Define a von Neumann entropy $S_{\mathrm{vN}}(AA^*:BB^*) = -\rho_{AA^*} \log \rho_{AA^*}$ with the reduced density matrix $\rho_{AA^*} = \mathrm{Tr}_{B B^*} \rho_{AA^* BB^*}$. Then the EoP is the optimization $E_P = \min_{A^* B^*}  S_{\mathrm{vN}}(AA^*:BB^*)$ over all  $A^*$ and $B^*$. 
In the holographic context, the bulk dual of EoP is hypothesized to be a minimal surface in AdS , known as the entanglement wedge cross section (EWCS or $E_W$). In AdS$_3$, the EWCS is a geodesic with finite length. Nevertheless, the optimization over purifications is practically not workable, though the EWCS in AdS$_3$ is easy to calculate. 
For this reason, a canonical purification is proposed. In this purification, the auxiliary state is a CPT copy of $\rho_{AB}$. The measure defined in this way is the reflected entropy \citep{Dutta:2019gen}. The holographic dual of the reflected entropy, in AdS$_3$, is a closed curve glued by two identical geodesics. 

\begin{figure}[h]
\includegraphics[width=0.8\textwidth]{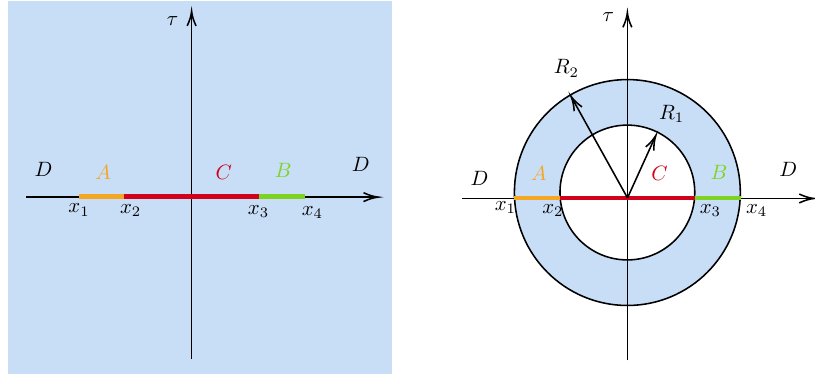}
\caption{Left panel: At $\tau=0$, two subsystems $A$ and $B$ on the $x$-axis
are in a mixed state $\rho_{AB}$, separated by segments $C$ and $D$. Right panel: After conformally subtracting $C$ and $D$ with two discs, an annulus is obtained, in which two subsystems $A$ and $B$ are now in a pure entangled state $\psi_{AB}$. \label{fig:alternative puri}}
\end{figure}

In  recent works \citep{Jiang:2024ijx, Jiang:2025tqu}, the SUBTRACTION approach is introduced to calculate the mixed state entanglement entropy in CFT successfully. Consider a pure system made up of nonoverlapping subsystems $A$, $B$, $C$ and $D$, where $A$ and $B$ are separated by $C$ and $D$, as shown in Figure \ref{fig:alternative puri}. Then $\rho_{AB}$ is a mixed state since $A$ and $B$ are not complementary.  Instead of adding auxiliary systems to replace $C$ and $D$ for purifying the mixed state $\rho_{AB}$, the subtraction approach simply removes subsystems $C$ and $D$ conformally. This operation makes $A$ and $B$ complementary. As a result, a pure state of $\psi_{AB}$ composed entirely of $A$ and $B$ is obtained. Then, calculating the entanglement entropy $S_{\mathrm{vN}}(A:B)$ between $A$ and $B$ is straightforward. 
In \citep{Jiang:2024ijx, Jiang:2025tqu}, the zero temperature infinite system is studied, for static and covariant scenarios.  It is confirmed  $S_{\mathrm{vN}}(A:B)=E_W (A:B)$. 

In this paper, we apply the subtraction approach to thermal CFT$_2$, whose holographic dual is the BTZ black hole.
The subtraction approach involves no optimization and thus 
provides a clear way to distinguish different configurations.
We will identify the holographic duals of the entanglement entropies more specifically.
We also study in thermofield double (TFD) state a two-sided mixed configuration slightly different from that addressed in \citep{Jiang:2024xqz}. We find that the  EWCS could cross the horizon at specific parameter values. 

The remainder of this paper is structured as follows:  In Section \ref{sec: CFT}, we  calculate mixed state entanglement entropy in thermal CFT$_2$ and address the holographic dual.  In section \ref{sec: TFD}, we study the thermofield double mixed state and find a horizon-crossing feature of the EWCS. Section \ref{sec: con} is the conclusion.

\section{Mixed state entanglement entropy in thermal CFT$_2$\label{sec: CFT}}

In this section, we first use the subtraction approach to calculate the entanglement entropies of mixed states in thermal CFT$_2$. Then, after identifying the holographic duals of the entanglement entropies respectively, we elucidate a delusive phase transition in the literature.

\subsection{Calculation in CFT}

We begin with the pure state $n$th R\'enyi entropy, defined as:
\begin{equation}\label{renyi}
S^{(n)}(A)=\frac{1}{1 - n}\mathrm{log}\mathrm{Tr}_A\rho^n_A,
\end{equation}
where $\rho_A =\mathrm{Tr}_{A^c}\, \rho_{tot} $ is the reduced density matrix after tracing out the complement of subsystem A over the total density matrix of the system. The $n$th power of the density matrix can be seen as the density matrix on an $n$-fold cover $\mathcal{M}_n$ of the original spacetime. This $\mathcal{M}_n$ is constructed by gluing $n$ copies of the sheet with a cut along $A$. The trace of $\rho^n_A$ is calculated using the partition function on $\mathcal{M}_n$:
\begin{equation}
\text{Tr}_A\rho_A^n=\frac{Z_n}{Z_1^n},
\label{eq: n copy}
\end{equation}
where the denominator $Z_1^n$ acts as a normalization.
The von Neumann entropy can be defined through the analytic continuation of R\'enyi entropy:
\begin{equation}\label{v}
S_{vN}(A)=\lim_{n\rightarrow 1}S^{(n)}(A)=-\text{Tr}_A\rho_A\text{log}\rho_A.
\end{equation}
So, once we find the partition functions of the original spacetime and the replicated manifold, we get the von Neumann entropy. This is the standard procedure to calculate the entanglement entropy for pure states.

The Kruskal extension of the BTZ black hole has two asymptotic boundaries, each associated with the spacetime boundary of the black hole. It is the holographic dual of the thermofield double state.

\begin{figure}[htb]
\includegraphics[width=0.6\textwidth]{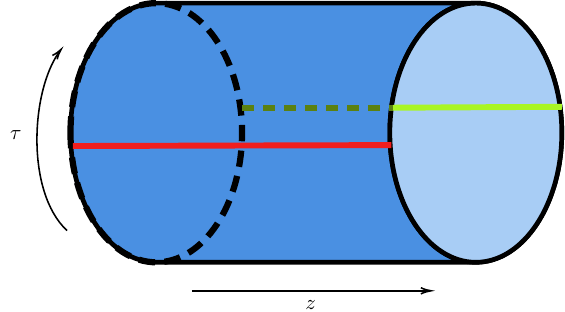}
\caption{$\rho = \ket{\text{TFD}}\bra{\text{TFD}}$ is prepared using Euclidean path integral. Two copies of CFTs reside on two slits, one on each side.}
    \label{fig: TFD}
\end{figure}

The thermofield double state $\ket{\text{TFD}}$ is a pure entangled state that describes a thermally excited system,  
\begin{equation}
	\ket{\text{TFD}}=\sum_{n}e^{-\frac{\beta}{2}E_n}\ket{n_L}\otimes\ket{n_R},
\end{equation}
where $\beta$ is the inverse temperature, $E_n$ is the energy of the $n$-th  eigenstate, and $\ket{n_L}$ and $\ket{n_R}$ are the states in the left and right CFTs, respectively.  
In this state, the two CFTs are entangled, with each CFT residing on one of the asymptotic boundaries of the BTZ black hole.
After tracing out one CFT (either $\ket{n_L}$ or $\ket{n_R}$), one gets the  density matrix of a thermal $\text{CFT}_2$,
\begin{equation}
	\rho_{thermal}=\frac{1}{\text{Tr}e^{-\beta H}}\sum_{n}\ket{n}\bra{n}e^{-\beta E_n}.
\end{equation}

In the Euclidean path integral formulation, the TFD state can be viewed as residing on an infinitely long cylinder, where the time direction is compactified and the two boundaries of the cylinder are connected by infinitely long slits, one on each side, as shown in Figure \ref{fig: TFD}. In this sense, the bulk region forms a solid cylinder.

\begin{figure}[h]
\includegraphics[width=0.6\textwidth]{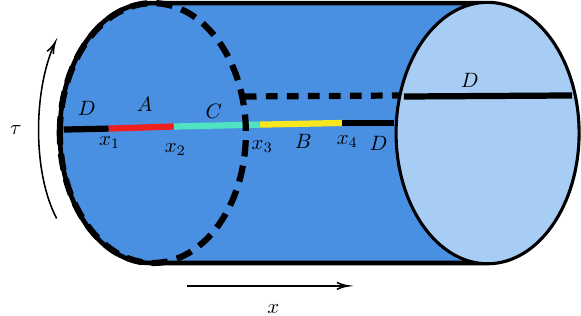}
\caption{At $\tau =0$, two  disjoint subsystems, $A=(x_1,x_2)$ and $B=(x_3,x_4)$, are noncomplementary, resulting in a mixed state $\rho_{AB}$. After tracing out $A$ and $B$,  $\rho_{CD}$ is also a mixed state.}
    \label{fig: ThermalCFT}
\end{figure}

Consider two subsystems $A=(x_1,x_2)$ and $B=(x_3,x_4)$, separated by segments $C$ and $D$,  at $\tau=0$,  as shown in Figure \ref{fig: ThermalCFT}. $\rho_{AB}$ obviously is a mixed state. We aim to study the entanglement between $A$ and $B$. The simplest way is to  map the cylinder onto the complex plane using the following conformal map:
\begin{equation}
	w=\text{exp}\left(\frac{2\pi z}{\beta}\right),
\end{equation}	
where $\beta$ is the inverse temperature of the thermal $\text{CFT}_2$ and is identical to the circumference of the cylinder. 

\begin{figure}[h]
\begin{centering}
\includegraphics[width=0.5\textwidth]{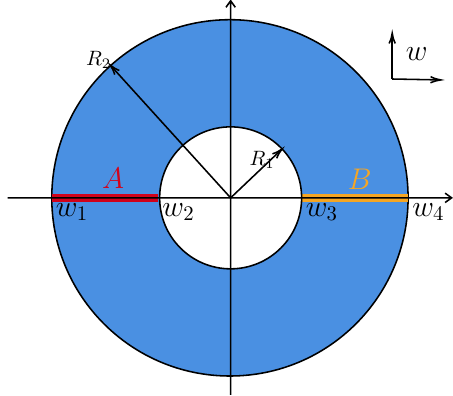}

\par\end{centering}    
	\caption{The subtraction approach is applied on the $w$ plane. Generally, segments $A=(w_1, w_2)$ and $B=(w_3, w_4)$ have different lengths. Two discs are used to remove segments $C=(w_2, w_3)$ and $D = (-\infty, w_1)\cup(w_4, \infty)$. The annulus with slits at $A$ and $B$ is clearly the density matrix of a pure state.}    
	\label{fig: annulus}
\end{figure}

To be self-contained, we follow the procedure introduced in refs \citep{Jiang:2024ijx, Jiang:2025tqu} to derive the entanglement between $A$ and $B$. On the $w$ plane, we use the subtraction approach to remove segments $C$ and $D$ by two discs with conformal boundary conditions, as shown in Figure \ref{fig: annulus}. These boundary conditions encode the information from $C$ and $D$. 
Generally, the resulting annulus is  asymmetric. Since every asymmetric annulus is conformally equivalent to a symmetric one through a global conformal transformation,  without loss of generality, we focus on the symmetric case and return to the asymmetric configuration at the end. The annulus with slits along $A$ and $B$ is clearly the density matrix of a pure state.
Thus, we can directly compute the entanglement between $A$ and $B$.

The partition function on a symmetric annulus is  \citep{Cardy:1989ir,Cardy:2004hm,Cardy:2016fqc}:
\begin{equation}
	Z_1=e^{cW/12}\sum_k\braket{a|k}\braket{k|b}e^{-2\delta_kW},
\end{equation}
where the conformal width $W$ is defined as $W = \log(R_2 / R_1)$, with the inner radius $R_1$ and outer radius $R_2$. Here, $c$ represents the central charge. $k$ indexes all allowed scalar operators within the annulus, each having a conformal dimension $\delta_k$. $\ket{a}$ and $\ket{b}$ are boundary states. They encode the information from the removed segments $C$ and $D$.

\begin{figure}[htb]
	\includegraphics[width=0.5\linewidth]{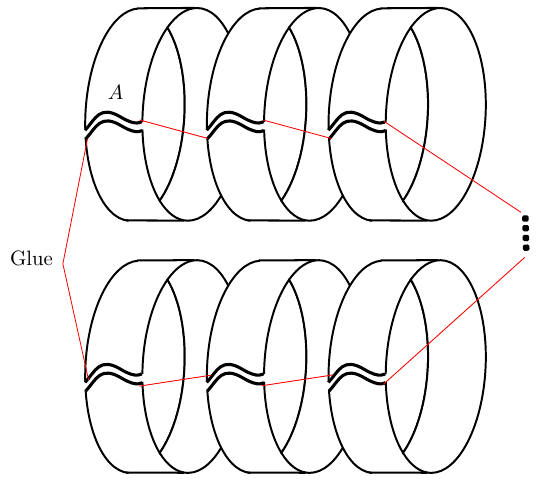}
	\centering
	\caption{The cut-and-glue procedure in the replica trick. Each annulus is cut along the segment $A$
and is glued with others cyclically. Red lines represent gluing operations.
Note that the resulted manifold is also an annulus.}
	\label{fig: replica}
\end{figure}

Next, we glue $n$ copies  of the original annulus along $A$ to get  $\mathcal{M}_n$ as shown in Figure \ref{fig: replica}. This construction yields  a manifold with a $\mathbb{Z}_n$ symmetry. Interestingly, the replicated manifold is conformally equivalent to the original annulus. Specifically, we can map $\mathcal{M}_n$ onto a cylinder via the coordinate transformation:
\begin{equation}
	y=n\log w.
\end{equation}
The Hamiltonian of $\mathcal{M}_n$ can be expressed using the translation generator on the cylinder:
\begin{equation}
	H=\int_0^{2\pi n}\frac{\text{d}\theta}{2\pi}\left(T(y)+\bar{T}(\bar{y})\right)=\frac{1}{n}(L_0+\bar{L}_0-\frac{c}{12}).
\end{equation}
We observe that the anomaly term we are interested in receives a contribution from the replica parameter in the denominator. This is equivalent to stating that the conformal width is rescaled to $W/n$. Therefore, the partition function of $\mathcal{M}_n$ is given by
\begin{equation}
	Z_n=e^{cW/12n}\sum_{k}\braket{a|k}\braket{k|b}e^{-2\delta_kW/n}.
\end{equation}
Since we are concerned with the vacuum entanglement, the contribution of excited states is irrelevant. We thus have

\begin{equation}\label{z}
	\begin{aligned}
	Z_1=&e^{cW/12}\braket{a|0}\braket{0|b},\\
	Z_n=&e^{cW/12n}\braket{a|0}\braket{0|b}.
	\end{aligned}
\end{equation}
Substituting eqn. \eqref{z} into eqns. \eqref{renyi} and \eqref{eq: n copy} yields the R\'enyi entropy:
\begin{equation}
	S^{(n)}(A)=\frac{c}{12}(1+\frac{1}{n})W+\text{log}(\braket{a|0}\braket{0|b}).
\end{equation}
The second term is the Affleck-Ludwig boundary
entropy \citep{Affleck:1991tk},  representing the information from the subtracted regions.   This boundary entropy is model dependent and are totally irrelevant to the entanglement between $A$ and $B$.  Therefore, 
the  entanglement entropy of $A$ and $B$ is:
\begin{equation}
	S_{vN}(A:B)=\lim_{n\rightarrow1}S^{(n)}(A) -\text{log}(\braket{a|0}\braket{0|b}) =\frac{c}{6}W.
\end{equation}

An easy way to get the entanglement entropy of the asymmetric configuration is to write the symmetric result in a conformal invariant manner. To this end, we rewrite $W=\log R_2/R_1$ in terms of the cross ratios $\zeta\equiv\frac{(w_2-w_1)(w_4-w_3)}{(w_3-w_2)(w_4-w_1)}$. Note that
\begin{equation}
	\frac{R_2}{R_1}=1+2\zeta+2\sqrt{\zeta(\zeta+1)}.
\end{equation}
Therefore, the static bipartite mixed state entanglement entropy in  thermal CFT$_2$ is
\begin{equation}
	S_{vN}(A:B)=\frac{c}{6}\text{log}\left(1+2\zeta+2\sqrt{\zeta(\zeta+1)}\right),
    \label{eq: MixedS-AB}
\end{equation}
where 
\begin{equation}
	\zeta=\frac{(w_2-w_1)(w_4-w_3)}{(w_3-w_2)(w_4-w_1)}=\frac{\text{sinh}(\pi(x_2-x_1)/\beta)\text{sinh}(\pi(x_4-x_3)/\beta)}{\text{sinh}(\pi(x_3-x_2)/\beta)\text{sinh}(\pi(x_4-x_1)/\beta)}.
    \label{eq: zeta}
\end{equation} 


Similarly, we can calculate the entanglement between $C=(x_2,x_3)$ and $D=(-\infty,x_1)\cup(x_4,\infty)$. The simplest way   is to make the  substitutions: $x_i\rightarrow x_{i+1}$ to get
\begin{equation}
	\zeta\to \zeta^{\prime}=\frac{(w_3-w_2)(w_1-w_4)}{(w_4-w_3)(w_1-w_2)}=\frac{1}{\zeta}.
\end{equation}
Therefore, the entanglement entropy for $C$ and $D$ is given by
\begin{equation}
	S_{vN}(C:D)=\frac{c}{6}\text{log}\left(1+\frac{2}{\zeta}+2\sqrt{\frac{1}{\zeta}\left(\frac{1}{\zeta}+1\right)}\right).
   \label{eq: MixedS-CD}
\end{equation}
The pure state entanglement entropy can be recovered by taking the limit:
\begin{equation}
	\lim_{\substack{x_2-x_1\to\epsilon\\x_4-x_3\to\epsilon}}S_{vN}(C:D)=\frac{c}{3}\text{log}\left(\frac{\beta}{\pi\epsilon}\text{sinh}\left(\frac{\pi \ell}{\beta}\right)\right),
\end{equation}
where  $\epsilon$ is the UV regulator and $\ell=x_3-x_2$ is the length of the interval.

As shown in ref. \citep{Jiang:2025tqu}, the covariant entanglement entropy which includes time dependence can be obtained by simply replacing $x_i$ with  $z_i=x_i+\tau_i$\footnote{Note in  \citep{Jiang:2025tqu}, the cross ratio used is $\eta = \frac{(w_2-w_1)(w_4-w_3)}{(w_3-w_1)(w_4-w_2)} = \frac{\zeta}{\zeta+1}$.}. 

It is worth noting that the competition between the two entanglement entropies, $S_{vN}(A:B)$ and $S_{vN}(C:D)$ leads to a phase transition at $\zeta=1$. This is guaranteed by the properties of hyperbolic geometry. We will see this more clearly in the holographic picture later. 

\subsection{Holographic computation}

In ref. \citep{Takayanagi:2017knl}, it was proposed that the holographic dual of  mixed state entanglement entropy 
is the EWCS. This has  been verified  in refs \citep{Jiang:2024ijx, Jiang:2025tqu} for zero temperature CFT. Now  we verify it for thermal CFT$_2$.


The EWCS  is a codimension-two extremal surface within the entanglement wedge that connects two boundary subregions. In the case of $\text{AdS}_3/\text{CFT}_2$, the EWCS is simply a geodesic. 
We work in the planar BTZ black hole, which is the holographic dual of the thermal $\text{CFT}_2$. The metric is given by
\begin{equation}
	ds^2=\frac{R_{AdS}^2}{z^2}\left(-(1-z^2/z_H^2)dt^2+(1-z^2/z_H^2)^{-1}dz^2+dx^2\right),
\end{equation}
where $z_H=\beta/2\pi$ denotes the horizon of the black hole and the coordinate range is $t,x\in\mathbb{R}$. 
The BTZ black hole can be understood as a quotient space of pure $\text{AdS}_3$ via the following coordinate transformation:
\begin{equation}
	\begin{aligned}\label{eq: trans}
		X&=\sqrt{1-z^2/z_H^2}\text{cosh}(t/z_H)e^{x/z_H},\\
		T&=\sqrt{1-z^2/z_H^2}\text{sinh}(t/z_H)e^{x/z_H},\\
		Z&=\frac{z}{z_H}e^{x/z_H}.
	\end{aligned}
\end{equation}
The new coordinates $\{X,T,Z\}$ correspond to a patch of Poincar\'e coordinates since the conditions $X>|T|$ and $Z<e^{x/z_H}$ are always satisfied. The region it covers in Poincar\'e patch is called the Rindler wedge.

\begin{figure}[htb]
	\begin{centering}
\includegraphics[width=0.5\textwidth]{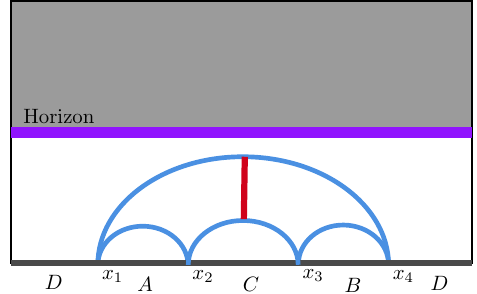}
    \end{centering}
	\caption{BTZ black hole in planar coordinates. The boundary is at $z=0$ and consists of $A=(x_1,x_2)$, $B=(x_3,x_4)$, $C=(x_2,x_3)$, and $D=(-\infty,x_1)\cup(x_4,\infty)$. The red line is the EWCS for $S_{vN}(A:B)$. The four RT surfaces, printed in blue, form an ideal quadrilateral within the entanglement wedge.}
	\label{fig: BTZ}
\end{figure}

Consider the configuration in Figure \ref{fig: BTZ}. The EWCS dual to $S_{vN}(A:B)$ shown in red is the minimal cross section in the entanglement wedge.

The length of the red line can be computed using the fact that the BTZ geometry is a quotient spacetime of pure $\text{AdS}_3$. In pure AdS$_3$, the geodesic length for a EWCS is\footnote{There are two methods to calculate $E_W(A:B)$. The simplest way is to use the ultraparallel  theorem which asserts that $E_W(A:B)$ is a unique geodesic. The second approach is to perform optimization as demonstrated in the Appendix \ref{app}.}:
\begin{equation}
	E_W(A:B)=\frac{c}{6}\text{log}\left(1+2\zeta+2\sqrt{\zeta(\zeta+1)}\right),
    \label{eq: ew}
\end{equation}
where the Brown-Henneaux formula $3R_{AdS}/2G^{(3)}$ \cite{Brown1986} is used, and $\zeta\equiv\frac{(X_2-X_1)(X_4-X_3)}{(X_3-X_2)(X_4-X_1)}$ is the cross ratio in terms of the pure AdS$_3$ parameters.

Using the coordinate transformation \eqref{eq: trans},    $E_W(A:B)$ in BTZ is still given by eqn. \eqref{eq: ew} but with  cross ratio in terms of the BTZ parameters:
\begin{equation}
	\zeta=\frac{\text{sinh}(\pi(x_2-x_1)/\beta)\text{sinh}(\pi(x_4-x_3)/\beta)}{\text{sinh}(\pi(x_3-x_2)/\beta)\text{sinh}(\pi(x_4-x_1)/\beta)}.
\end{equation}
Thus, comparing with eqn.  \eqref{eq: MixedS-AB},  we see:
\begin{equation}
	S_{vN}(A:B)=E_W(A:B).
\end{equation}	
Again, by setting $\zeta\rightarrow1/\zeta$, we confirm
\begin{equation}
	E_W(C:D)=S_{vN}(C:D)=\frac{c}{6}\log\left(1+\frac{2}{\zeta}+2\sqrt{\frac{1}{\zeta}\left(\frac{1}{\zeta}+1\right)}\right).
\end{equation}

\subsection{The phase transition}

In pure AdS$_3$,  it is easy to distinguish the EWCS associated with different segments. However, it might be tricky in other asymptotic AdS geometries, such as BTZ. To  identify the EWCS dual to a specific segment, the typical process is to find all possible  surfaces and then pick the minimal one. This is in line with the optimization procedure in the EoP definition. Nevertheless, the subtraction approach in CFT involves no optimization and has no ambiguity. The entanglement entropy obtained in CFT this way is unique. So, it is a more accurate approach for making distinctions.

\begin{figure}[htb]
	\begin{centering}
\includegraphics[width=0.5\textwidth]{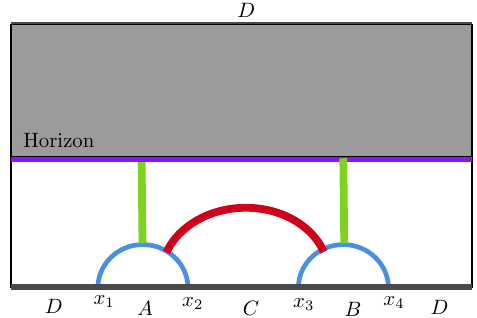}
    \end{centering}
	\caption{The red line is $E_W(C:D)$. The LHS green line is $E_W\Big((-\infty,x_1):(x_2,\infty)\Big)$. The RHS green line is $E_W\Big((-\infty,x_3):(x_4,\infty)\Big)$}
	\label{fig: EWBTZ}
\end{figure}

To make the discussion concrete, let us consider the configuration shown in Figure \ref{fig: EWBTZ}. 
Solely from the bulk EWCS investigation, it appears that 
there are two candidates for $E_W(C:D)$,  the combination of the two green straight lines  
or the red line. 
However, we are going to show that the LHS (RHS) green line should be understood as the EWCS for subsystems $(-\infty,x_1)$ and $(x_2,\infty)$ ($(-\infty,x_3)$ and $(x_4,\infty)$), rather than part of $E_W(C:D)$.

The length of the green line on the left is
\begin{equation}
	L_{green}=\int^{z_H}_{z^{*}}\frac{dz}{z\sqrt{1-z^2/z_H^2}}=\text{log}\left(\text{coth}\left(\frac{\pi(x_2-x_1)}{2\beta}\right)\right),
\end{equation}
where the turning point is given by $z^*=z_H\text{tanh}(\ell/2z_H)$ for a subsystem of size $\ell$. 
On the other hand, referring to eqns. \eqref{eq: MixedS-AB}, \eqref{eq: zeta} and Figure \ref{fig: BTZ},  under the limits: $x_1\rightarrow-\infty$, $x_4\rightarrow\infty$,  the entanglement entropy $S_{vN}(A:B)$ is
\begin{equation}
    \lim_{\substack{x_1\to -\infty\\x_4\to +\infty}}\, \frac{6}{c}\, S_{vN}(A:B)\\
		=\text{log}\left(\text{coth}\left(\frac{\pi (x_3 -x_2)}{2\beta}\right)\right).
\end{equation}	
So, the LHS (RHS) green line in Figure \ref{fig: EWBTZ} really is the EWCS for $(-\infty,x_1)$ and $(x_2,\infty)$ ($(-\infty,x_3)$ and $(x_4,\infty)$).

\section{Two-sided entanglement configuration\label{sec: TFD}}

\begin{figure}[htb]
\includegraphics[width=0.5\textwidth]{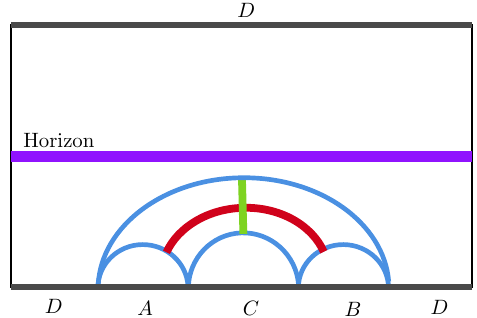}

	\caption{A slice of the solid cylinder. Last section discussion is an one-sided entanglement. In this configuration, the ideal quadrilateral in entanglement wedge, depicted in blue, will not touch the horizon. The EWCSs for $S_{vN}(A:B)$ and $S_{vN}(C:D)$ are shown in green and red, respectively.}
	\label{fig: OneSide}
\end{figure}

In the last section, we primarily focused on the entanglement within the same side, as shown in Figure \ref{fig: OneSide}.
This one-sided entanglement configuration is the usual one considered in thermal CFT. 
However, examining entanglement across different sides is quite insightful. For example $A\in\ket{n_L}$ while $B\in\ket{n_R}$.  One such well-designed configuration has been studied in \citep{Jiang:2024xqz} to realize ER=EPR successfully.

\begin{figure}[htb]
\includegraphics[width=0.5\textwidth]{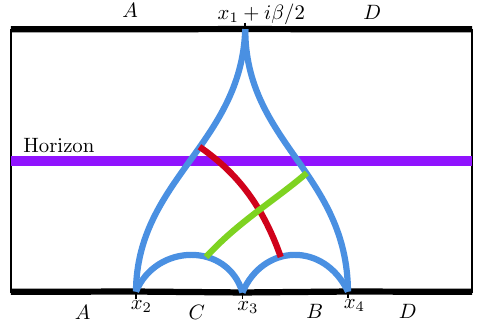}
	\caption{A two-sided entanglement configuration. The ideal quadrilateral within the entanglement wedge is printed in blue. The EWCS in red color corresponding to $S_{vN}(C:D)$ could cross the horizon.}
	\label{fig: TwoSide}
\end{figure}

We thus turn  to a configuration shown in Figure \ref{fig: TwoSide}. In this setup, the entanglement entropy can be directly read off from eqns. \eqref{eq: MixedS-AB}  and \eqref{eq: MixedS-CD}:
\begin{equation}
	\begin{aligned}
		S_{vN}(A:B)&=\frac{c}{6}\text{log}\left(1+2\zeta+2\sqrt{\zeta(\zeta+1)}\right)\\
		S_{vN}(C:D)&=\frac{c}{6}\text{log}\left(1+\frac{2}{\zeta}+2\sqrt{\frac{1}{\zeta}\left(\frac{1}{\zeta}+1\right)}\right),
	\end{aligned}
\end{equation}
where 
\begin{equation}
	\zeta=\frac{\text{cosh}(\pi(x_2-x_1)/\beta)\text{sinh}(\pi(x_4-x_3)/\beta)}{\text{cosh}(\pi(x_4-x_1)/\beta)\text{sinh}(\pi(x_3-x_2)/\beta)}.
\end{equation}
is the modified cross ratio.

In contrast to the one-sided entanglement, two-sided entanglement has a distinct feature:  the dual EWCS could pass through the horizon, as shown in Figure \ref{fig: TwoSide}.  An easy way to determine the horizon-crossing point is to map the BTZ geometry onto pure $\text{AdS}_3$ by using the  coordinate transformation \eqref{eq: trans}.
With this transformation, the horizon is mapped to the vertical line passing through the origin. 
Therefore,  once the horizontal coordinate of the endpoint of the EWCS equals zero, the  EWCS crosses  the horizon.

In pure $\text{AdS}_3$, with the results derived in the Appendix \ref{app}, the endpoints of $E_W(C:D)$ are given by\footnote{We have slightly adjusted the notation compared to the Appendix. Here, the capital letters represent the Poincar\'e coordinates in the bulk, while $w_i$ represent the coordinates on the boundary.}:
\begin{equation}\label{eq: TFD-Trans}
	\begin{aligned}
		X_L&=\frac{w_2(w_3-w_1)(w_4-w_1)+w_1(w_3-w_2)(w_4-w_2)}{(w_3-w_1)(w_4-w_1)+(w_3-w_2)(w_4-w_2)},\\
		Z_{L}&=\frac{(w_2-w_1)\sqrt{(w_3-w_2)(w_3-w_1)(w_4-w_2)(w_4-w_1)}}{(w_3-w_1)(w_4-w_1)+(w_3-w_2)(w_4-w_2)},\\
		X_R&=\frac{w_3(w_4-w_2)(w_4-w_1)+w_4(w_3-w_2)(w_3-w_1)}{(w_4-w_2)(w_4-w_1)+(w_3-w_2)(w_3-w_1)},\\
		Z_R&=\frac{(w_4-w_3)\sqrt{(w_3-w_2)(w_3-w_1)(w_4-w_2)(w_4-w_1)}}{(w_4-w_2)(w_4-w_1)+(w_3-w_2)(w_3-w_1)},
	\end{aligned}
\end{equation}
where $(X_L,Z_L)$ and $(X_R,Z_R)$ represent the coordinates of the left and right endpoints of $E_W(C:D)$, respectively. To align with our previous discussion, we use $w$ to denote the coordinates on the flat boundary of Poincar\'e half plane. The relation between $w$ and $x$ is given by
\begin{equation}
	w=\text{exp}\left(\frac{2\pi x}{\beta}\right).
    \label{eq: TFD-Map}
\end{equation}
The horizon-crossing occurs at $X_L=0$. This leads to:

\begin{equation}
\frac{e^{\frac{2\pi x_2}{\beta}}(e^{\frac{2\pi x_3}{\beta}}+e^{\frac{2\pi x_1}{\beta}})(e^{\frac{2\pi x_4}{\beta}}+e^{\frac{2\pi x_1}{\beta}})-e^{\frac{2\pi x_1}{\beta}}(e^{\frac{2\pi x_3}{\beta}}-e^{\frac{2\pi x_2}{\beta}})(e^{\frac{2\pi x_4}{\beta}}-e^{\frac{2\pi x_2}{\beta}})}{(e^{\frac{2\pi x_3}{\beta}}+e^{\frac{2\pi x_1}{\beta}})(e^{\frac{2\pi x_4}{\beta}}+e^{\frac{2\pi x_1}{\beta}})+(e^{\frac{2\pi x_3}{\beta}}-e^{\frac{2\pi x_2}{\beta}})(e^{\frac{2\pi x_4}{\beta}}-e^{\frac{2\pi x_2}{\beta}})}=0,
\end{equation}
which can be further simplified to:
\begin{equation}\label{k}
	\kappa\equiv\frac{\text{cosh}(\pi(x_3-x_1)/\beta)\text{cosh}(\pi(x_4-x_1)/\beta)}{\text{sinh}(\pi(x_3-x_2)/\beta)\text{sinh}(\pi(x_4-x_2)/\beta)}=1.	
\end{equation}	
Thus, the behavior of $E_W(C:D)$ in the extended BTZ black hole depends on $\kappa$:
\begin{itemize}
    \item If $\kappa>1$, $E_W(C:D)$ does not cross the horizon.
    \item If $\kappa<1$, $E_W(C:D)$ crosses the horizon.
\end{itemize}
At the horizon-crossing point, the cross ratio simplifies to: \begin{equation}
	\zeta^{*}=\left(\frac{\text{cosh}(\pi(x_3-x_1)/\beta)}{\text{sinh}(\pi(x_3-x_2)/\beta)}\right)^2-1=\left(\frac{\text{sinh}(\pi(x_4-x_2)/\beta)}{\text{cosh}(\pi(x_4-x_1)/\beta)}\right)^2-1.	
\end{equation}	

The same analysis can be applied to $S_{vN}(A:B)$, depicted in green in Figure \ref{fig: TwoSide}. By setting the horizontal coordinate of the endpoint of $E_W(A:B)$ equal to zero, we obtain
\begin{equation}\label{k2}
     \bar{\kappa}\equiv\frac{\text{cosh}(\pi(x_2-x_1)/\beta)\text{cosh}(\pi(x_3-x_1)/\beta)}{\text{sinh}(\pi(x_4-x_2)/\beta)\text{sinh}(\pi(x_4-x_3)/\beta)}=1.	
\end{equation}
Thus, the behavior of $E_W(A:B)$ in the extended BTZ black hole depends on $\bar{\kappa}$:
\begin{itemize}
	\item If $\bar{\kappa}>1$, $E_W(A:B)$ does not cross the horizon.
	\item If $\bar{\kappa}<1$, $E_W(A:B)$ crosses the horizon.
\end{itemize}
At the  horizon-crossing point, the cross ratio simplifies to:
\begin{equation}
	\zeta^*=\left(\left(\frac{\text{cosh}(\pi(x_3-x_1)/\beta)}{\text{sinh}(\pi(x_4-x_3)/\beta)}\right)^2-1\right)^{-1}=\left(\left(\frac{\text{sinh}(\pi(x_4-x_2)/\beta)}{\text{cosh}(\pi(x_2-x_1)/\beta)}\right)^2-1\right)^{-1}.
\end{equation}

\section{Conclusion\label{sec: con}}
In this paper, we employed the subtraction approach to study the bipartite mixed state entanglement entropy of thermal $\text{CFT}_2$.  In planar BTZ black hole, our analysis provided a clear way to identify the holographic duals of different entanglement entropies unambiguously.
Additionally, we investigated bipartite mixed state entanglement in the TFD state. We demonstrated that, at certain parameter values, the EWCS can extend across the horizon. 

\vspace*{3.0ex}
\begin{acknowledgments}
\paragraph*{Acknowledgments.} 
This work is supported in part by NSFC (Grant No.12275184).
\end{acknowledgments}



\bibliographystyle{unsrturl}
\bibliography{ref202504}	

\begin{thebibliography}{10}

\bibitem{Maldacena:1997re}
Juan~Martin Maldacena.
\newblock {The Large N limit of superconformal field theories and
  supergravity}.
\newblock {\em Adv. Theor. Math. Phys.}, 2:231--252, 1998.
\newblock \href {http://arxiv.org/abs/hep-th/9711200}
  {\path{arXiv:hep-th/9711200}}, \href
  {http://dx.doi.org/10.1023/A:1026654312961}
  {\path{doi:10.1023/A:1026654312961}}.

\bibitem{Witten:1998qj}
Edward Witten.
\newblock {Anti-de Sitter space and holography}.
\newblock {\em Adv. Theor. Math. Phys.}, 2:253--291, 1998.
\newblock \href {http://arxiv.org/abs/hep-th/9802150}
  {\path{arXiv:hep-th/9802150}}, \href
  {http://dx.doi.org/10.4310/ATMP.1998.v2.n2.a2}
  {\path{doi:10.4310/ATMP.1998.v2.n2.a2}}.

\bibitem{Gubser:1998bc}
S.~S. Gubser, Igor~R. Klebanov, and Alexander~M. Polyakov.
\newblock {Gauge theory correlators from noncritical string theory}.
\newblock {\em Phys. Lett. B}, 428:105--114, 1998.
\newblock \href {http://arxiv.org/abs/hep-th/9802109}
  {\path{arXiv:hep-th/9802109}}, \href
  {http://dx.doi.org/10.1016/S0370-2693(98)00377-3}
  {\path{doi:10.1016/S0370-2693(98)00377-3}}.

\bibitem{Ryu:2006bv}
Shinsei Ryu and Tadashi Takayanagi.
\newblock {Holographic derivation of entanglement entropy from AdS/CFT}.
\newblock {\em Phys. Rev. Lett.}, 96:181602, 2006.
\newblock \href {http://arxiv.org/abs/hep-th/0603001}
  {\path{arXiv:hep-th/0603001}}, \href
  {http://dx.doi.org/10.1103/PhysRevLett.96.181602}
  {\path{doi:10.1103/PhysRevLett.96.181602}}.

\bibitem{Ryu:2006ef}
Shinsei Ryu and Tadashi Takayanagi.
\newblock Aspects of holographic entanglement entropy.
\newblock {\em Journal of High Energy Physics}, 2006(08):045--045, aug 2006.
\newblock \href {http://dx.doi.org/10.1088/1126-6708/2006/08/045}
  {\path{doi:10.1088/1126-6708/2006/08/045}}.

\bibitem{Hubeny:2007xt}
Veronika~E. Hubeny, Mukund Rangamani, and Tadashi Takayanagi.
\newblock {A Covariant holographic entanglement entropy proposal}.
\newblock {\em JHEP}, 07:062, 2007.
\newblock \href {http://arxiv.org/abs/0705.0016} {\path{arXiv:0705.0016}},
  \href {http://dx.doi.org/10.1088/1126-6708/2007/07/062}
  {\path{doi:10.1088/1126-6708/2007/07/062}}.

\bibitem{VanRaamsdonk:2010pw}
Mark Van~Raamsdonk.
\newblock {Building up spacetime with quantum entanglement}.
\newblock {\em Gen. Rel. Grav.}, 42:2323--2329, 2010.
\newblock \href {http://arxiv.org/abs/1005.3035} {\path{arXiv:1005.3035}},
  \href {http://dx.doi.org/10.1142/S0218271810018529}
  {\path{doi:10.1142/S0218271810018529}}.

\bibitem{Jiang:2024xcy}
Xin Jiang, Peng Wang, Houwen Wu, and Haitang Yang.
\newblock {Einstein Equation Governs the Dynamics of Entanglement Entropy in
  CFT}.
\newblock 10 2024.
\newblock \href {http://arxiv.org/abs/2410.19711} {\path{arXiv:2410.19711}}.

\bibitem{Jiang:2024xqz}
Xin Jiang, Peng Wang, Houwen Wu, and Haitang Yang.
\newblock {Realization of ''ER=EPR''}.
\newblock 11 2024.
\newblock \href {http://arxiv.org/abs/2411.18485} {\path{arXiv:2411.18485}}.

\bibitem{Calabrese:2012nk}
Pasquale Calabrese, John Cardy, and Erik Tonni.
\newblock {Entanglement negativity in extended systems: A field theoretical
  approach}.
\newblock {\em J. Stat. Mech.}, 1302:P02008, 2013.
\newblock \href {http://arxiv.org/abs/1210.5359} {\path{arXiv:1210.5359}},
  \href {http://dx.doi.org/10.1088/1742-5468/2013/02/P02008}
  {\path{doi:10.1088/1742-5468/2013/02/P02008}}.

\bibitem{Takayanagi:2017knl}
Tadashi Takayanagi and Koji Umemoto.
\newblock {Entanglement of purification through holographic duality}.
\newblock {\em Nature Phys.}, 14(6):573--577, 2018.
\newblock \href {http://arxiv.org/abs/1708.09393} {\path{arXiv:1708.09393}},
  \href {http://dx.doi.org/10.1038/s41567-018-0075-2}
  {\path{doi:10.1038/s41567-018-0075-2}}.

\bibitem{Tamaoka:2018ned}
Kotaro Tamaoka.
\newblock {Entanglement Wedge Cross Section from the Dual Density Matrix}.
\newblock {\em Phys. Rev. Lett.}, 122(14):141601, 2019.
\newblock \href {http://arxiv.org/abs/1809.09109} {\path{arXiv:1809.09109}},
  \href {http://dx.doi.org/10.1103/PhysRevLett.122.141601}
  {\path{doi:10.1103/PhysRevLett.122.141601}}.

\bibitem{Dutta:2019gen}
Souvik Dutta and Thomas Faulkner.
\newblock {A canonical purification for the entanglement wedge cross-section}.
\newblock {\em JHEP}, 03:178, 2021.
\newblock \href {http://arxiv.org/abs/1905.00577} {\path{arXiv:1905.00577}},
  \href {http://dx.doi.org/10.1007/JHEP03(2021)178}
  {\path{doi:10.1007/JHEP03(2021)178}}.

\bibitem{Wen:2021qgx}
Qiang Wen.
\newblock {Balanced Partial Entanglement and the Entanglement Wedge Cross
  Section}.
\newblock {\em JHEP}, 04:301, 2021.
\newblock \href {http://arxiv.org/abs/2103.00415} {\path{arXiv:2103.00415}},
  \href {http://dx.doi.org/10.1007/JHEP04(2021)301}
  {\path{doi:10.1007/JHEP04(2021)301}}.

\bibitem{Mori:2024gwe}
Takato Mori and Beni Yoshida.
\newblock {Does connected wedge imply distillable entanglement?}
\newblock 11 2024.
\newblock \href {http://arxiv.org/abs/2411.03426} {\path{arXiv:2411.03426}}.

\bibitem{Jiang:2024ijx}
Xin Jiang, Peng Wang, Houwen Wu, and Haitang Yang.
\newblock {Alternative to purification in conformal field theory}.
\newblock {\em Phys. Rev. D}, 111(2):L021902, 2025.
\newblock \href {http://arxiv.org/abs/2406.09033} {\path{arXiv:2406.09033}},
  \href {http://dx.doi.org/10.1103/PhysRevD.111.L021902}
  {\path{doi:10.1103/PhysRevD.111.L021902}}.

\bibitem{Jiang:2025tqu}
Xin Jiang, Houwen Wu, Peng Wang, and Haitang Yang.
\newblock {Mixed State Entanglement Entropy in CFT}.
\newblock 1 2025.
\newblock \href {http://arxiv.org/abs/2501.08198} {\path{arXiv:2501.08198}}.

\bibitem{Cardy:1989ir}
John~L. Cardy.
\newblock {Boundary Conditions, Fusion Rules and the Verlinde Formula}.
\newblock {\em Nucl. Phys. B}, 324:581--596, 1989.
\newblock \href {http://dx.doi.org/10.1016/0550-3213(89)90521-X}
  {\path{doi:10.1016/0550-3213(89)90521-X}}.

\bibitem{Cardy:2004hm}
John~L. Cardy.
\newblock {Boundary conformal field theory}.
\newblock 11 2004.
\newblock \href {http://arxiv.org/abs/hep-th/0411189}
  {\path{arXiv:hep-th/0411189}}.

\bibitem{Cardy:2016fqc}
John Cardy and Erik Tonni.
\newblock {Entanglement hamiltonians in two-dimensional conformal field
  theory}.
\newblock {\em J. Stat. Mech.}, 1612(12):123103, 2016.
\newblock \href {http://arxiv.org/abs/1608.01283} {\path{arXiv:1608.01283}},
  \href {http://dx.doi.org/10.1088/1742-5468/2016/12/123103}
  {\path{doi:10.1088/1742-5468/2016/12/123103}}.

\bibitem{Affleck:1991tk}
Ian Affleck and Andreas W.~W. Ludwig.
\newblock {Universal noninteger 'ground state degeneracy' in critical quantum
  systems}.
\newblock {\em Phys. Rev. Lett.}, 67:161--164, 1991.
\newblock \href {http://dx.doi.org/10.1103/PhysRevLett.67.161}
  {\path{doi:10.1103/PhysRevLett.67.161}}.

\bibitem{Brown1986}
J.~David Brown and M.~Henneaux.
\newblock Central charges in the canonical realization of asymptotic
  symmetries: An example from three-dimensional gravity.
\newblock {\em Commun. Math. Phys.}, 104:207--226, 1986.
\newblock \href {http://dx.doi.org/10.1007/BF01211590}
  {\path{doi:10.1007/BF01211590}}.

\end{thebibliography}

\appendix
\section{Endpoints of EWCS in Pure $\text{AdS}_3$}
\label{app}

We work in pure $\text{AdS}_3$, where the metric in Poincar\'e coordinates is given by
\begin{equation}
	ds^2=\frac{-dt^2+dx^2+dz^2}{z^2}.
\end{equation}
The EWCS is defined as the minimal cross-sectional area of the entanglement wedge. It is uniquely determined once the boundary points are specified. In this Appendix, our goal is to derive the explicit formulae for the lengths and endpoints of the EWCSs in terms of the boundary coordinates.

\begin{figure}[htb]
\includegraphics[width=0.5\textwidth]{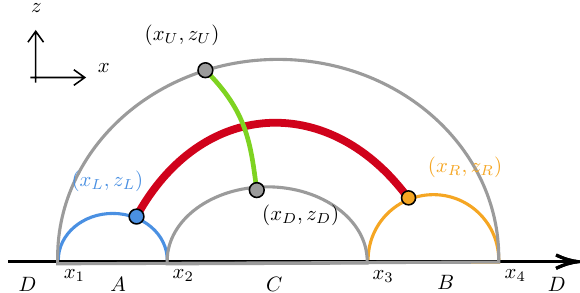}
	\caption{EWCS in Poincar\'e coordinate. The endpoints, $(x_L,z_L)$ and $(x_R,z_R)$, can move freely along the RT surfaces of $A$ and $B$, respectively. The endpoints of the EWCS are determined through an optimization procedure, where the distance between $(x_L,z_L)$ and $(x_R,z_R)$ is extremized.}
	\label{saddle}
\end{figure}

We restrict our discussion to the $t=0$ slice. Consider the bipartite entanglement between two subsystems, $C=(x_2,x_3)$ and $D=(-\infty,x_1)\cup(x_4,\infty)$ located on the boundary, as illustrated in Figure (\ref{saddle}). We can calculate its length and endpoints by the optimization procedure.

The RT surfaces of $A$ and $B$ can be parametrized as:
\begin{equation}\label{ep}
		\begin{aligned}
		x_L(u)&=\frac{x_1+x_2}{2}+\frac{x_2-x_1}{2}u,\\
		z_L(u)&=\frac{x_2-x_1}{2}\sqrt{1-u^2},\\
		x_R(v)&=\frac{x_3+x_4}{2}+\frac{x_4-x_3}{2}v,\\
		z_R(v)&=\frac{x_4-x_3}{2}\sqrt{1-v^2},
	\end{aligned}
\end{equation}
where we have used the fact that each geodesic in the Poincar\'e coordinate is a segment of a semicircle. Here, $u=\text{cos}\theta_A$ and $v=\text{cos}\theta_B$ both take values in the interval $(0,1)$.

The distance between two given points, $(x_L,z_L)$ and $(x_R,z_R)$, is given by:
\begin{equation}
	d_H(x_L,z_L,x_R,z_R)=\text{arccosh}\left(1+\frac{(x_R-x_L)^2+(z_R-z_L)^2}{2z_Lz_R}\right).
\end{equation}
We then apply the saddle condition, which involves setting the derivatives of the distance with respect to $u$ and $v$ to zero:
\begin{equation}
		\frac{\text{d}d_H(u,v)}{\text{d}u}=0\quad\text{and}\quad
		\frac{\text{d}d_H(u,v)}{\text{d}v}=0.
\end{equation}
After simplification, we arrive at two quadratic equations:
\begin{equation}
	\begin{aligned}
		&(x_4-x_3)(x_4+x_3-x_2-x_1)uv-((x_3-x_1)^2+(x_4-x_2)^2+(x_2-x_1)(x_4-x_3))u\\
		&-(x_2-x_1)(x_4-x_3)v+(x_2-x_1)(x_4+x_3-x_2-x_1)=0\\
		&(x_2-x_1)(x_4+x_3-x_2-x_1)uv-((x_3-x_1)^2+(x_4-x_2)^2+(x_2-x_1)(x_4-x_3))v\\
		&-(x_2-x_1)(x_4-x_3)u+(x_4-x_3)(x_4+x_3-x_2-x_1)=0.
	\end{aligned}
\end{equation}
Solving these quadratic equations yields:
\begin{equation}
	u=\left(1+\frac{2(x_3-x_2)(x_4-x_2)}{(x_2-x_1)(x_4+x_3-x_2-x_1)}\right)^{-1},\quad\quad
	v=\left(1+\frac{2(x_3-x_2)(x_3-x_1)}{(x_4-x_3)(x_4+x_3-x_2-x_1)}\right)^{-1}.
\end{equation}
Another solution has been dropped out, since $u\in(0,1)$ and $v\in(0,1)$. Substituting these values into \eqref{ep}, we arrive at the final result for the endpoints of $E_W(C:D)$,
\begin{equation}
	\begin{aligned}
		x_L&=\frac{x_2(x_3-x_1)(x_4-x_1)+x_1(x_3-x_2)(x_4-x_2)}{(x_3-x_1)(x_4-x_1)+(x_3-x_2)(x_4-x_2)},\\
		z_{L}&=\frac{(x_2-x_1)\sqrt{(x_3-x_2)(x_3-x_1)(x_4-x_2)(x_4-x_1)}}{(x_3-x_1)(x_4-x_1)+(x_3-x_2)(x_4-x_2)},\\
		x_R&=\frac{x_3(x_4-x_2)(x_4-x_1)+x_4(x_3-x_2)(x_3-x_1)}{(x_4-x_2)(x_4-x_1)+(x_3-x_2)(x_3-x_1)},\\
		z_R&=\frac{(x_4-x_3)\sqrt{(x_3-x_2)(x_3-x_1)(x_4-x_2)(x_4-x_1)}}{(x_4-x_2)(x_4-x_1)+(x_3-x_2)(x_3-x_1)}.
	\end{aligned}
\end{equation}
One can verify the expression of the length of $E_W(C:D)$ given by $d_H(x_1,x_2,x_3,x_4)$ can be simplified into a very concise form:
\begin{equation}
	E_W(C:D)=\frac{c}{6}\text{arccosh}\left(1+2\frac{(x_3-x_2)(x_4-x_1)}{(x_2-x_1)(x_4-x_3)}\right).
\end{equation}
Alternatively, in terms of the cross ratio $\zeta=\frac{(x_2-x_1)(x_4-x_3)}{(x_3-x_2)(x_4-x_1)}$, the expression becomes:
\begin{equation}
	E_W(C:D)=\frac{c}{6}\text{arccosh}\left(1+\frac{2}{\zeta}\right)=\frac{c}{6}\text{log}\left(1+\frac{2}{\zeta}+2\sqrt{\frac{1}{\zeta}\left(\frac{1}{\zeta}+1\right)}\right),
\end{equation}
which matches the CFT calculation.

The similar analysis can be applied to the EWCS for $S_{vN}(A:B)$. The results are given by
\begin{equation}
	\begin{aligned}
		x_U=&\frac{x_4(x_2-x_1)(x_3-x_1)+x_1(x_4-x_2)(x_4-x_3)}{(x_2-x_1)(x_3-x_1)+(x_4-x_2)(x_4-x_3)},\\
		z_U=&\frac{(x_4-x_1)\sqrt{(x_2-x_1)(x_3-x_1)(x_4-x_2)(x_4-x_3)}}{(x_2-x_1)(x_3-x_1)+(x_4-x_2)(x_4-x_3)},\\
		x_D=&\frac{x_2(x_4-x_3)(x_3-x_1)+x_3(x_2-x_1)(x_4-x_2)}{(x_4-x_3)(x_3-x_1)+(x_2-x_1)(x_4-x_2)},\\
		z_D=&\frac{(x_3-x_2)\sqrt{(x_3-x_4)(x_3-x_1)(x_2-x_1)(x_2-x_4)}}{(x_4-x_3)(x_3-x_1)+(x_2-x_1)(x_4-x_2)},
	\end{aligned}
\end{equation}
where the subscripts, $U$ and $D$, represent ``up'' and ``down'', respectively. The distance between $(x_U,z_U)$ and $(x_D,z_D)$ is given by
\begin{equation}
	E_W(A:B)=\frac{c}{6}\text{log}\left(1+2\zeta+2\sqrt{\zeta(\zeta+1)}\right).
\end{equation}

\end{document}